\documentclass[aps,%
reprint,
prl,
groupedaddress]{revtex4-2}
\usepackage{xcolor}
\usepackage{float}
\usepackage{amsmath,amsfonts,amssymb,epsfig,graphicx}
\usepackage{slashed} 
\usepackage{hyperref}
\usepackage[normalem]{ulem}
\graphicspath{{./pic/}}
\begin{document}

\title{Low Energy Neutrino and Mass Dark Matter Detection Using Freely Falling Atoms}

\author{Alim \surname{Ruzi}}
\email[]{alim.ruzi@pku.edu.cn}
\affiliation{State Key Laboratory of Nuclear Physics and Technology, School of Physics, Peking University, Beijing, 100871, China}

\author{Sitian \surname{Qian}}
\email[]{stqian@pku.edu.cn}
\affiliation{State Key Laboratory of Nuclear Physics and Technology, School of Physics, Peking University, Beijing, 100871, China}

\author{Tianyi \surname{Yang}}
\email[]{tyyang99@pku.edu.cn}
\affiliation{State Key Laboratory of Nuclear Physics and Technology, School of Physics, Peking University, Beijing, 100871, China}

\author{Qiang \surname{Li}}
\email[]{qliphy0@pku.edu.cn}
\affiliation{State Key Laboratory of Nuclear Physics and Technology, School of Physics, Peking University, Beijing, 100871, China}

\begin{abstract}
We propose a new method to detect low-energy neutrinos and low-mass dark matter at or below the MeV scale, through their coherent scatterings from freely falling heavy atoms and the resulting kinematic shifts. We start with a simple calculation for illustration: for $10^7$ heavy atoms of a mass number around 100 with a small recoil energy of 1 meV, the corresponding velocities can reach $0.01 {\rm m/s}$ and produce significant kinematic shifts that can be detected. We then show that the proposed device should be able to probe vast low-energy regions of neutrinos from meV to MeV and can surpass previous limits on sub-MeV dark matter by several orders of magnitude. Such a proposal can be useful to (1) detect sub-MeV-scale dark matter: with $10^2$ atom guns shooting downwards, for example, CsI or lead clusters consisting of $10^{7}$ atoms with a frequency around $10^3$ Hz, it can already be sensitive to scattering cross-sections at the level of $10^{-33 (-34)}\rm{cm}^{2}$ for 1 (0.1) MeV dark matter and surpass current limits. Technological challenges include high-quality atom cluster production and injections. (2) Measure coherent neutrino-nuclei scatterings at the 0.1-1 MeV region for the first time: with $10^4$ atom guns shooting downwards CsI clusters consisting of $10^{11}$ atoms and a frequency of $10^{6}$ Hz. One can expect 10 events from MeV solar neutrinos to be observed per year. Furthermore, (3) this method can be extended to probe very low-energy neutrinos down to the eV-KeV region and may be able to detect the cosmic neutrino background, although it remains challenging.
\end{abstract}

\maketitle

\section{Motivations} 
Low-energy neutrinos at or below the MeV scale are abundant on Earth. For example, the neutrino flux is dominated in the keV to MeV energy range by the neutrinos produced in the Sun~\cite{Vitagliano:2017ona}, and has been extensively measured recently by the Borexino experiment~\cite{borexino}, with an energy threshold of 0.19 MeV, by means of their weak elastic scattering off electrons. In the meV to 100 eV region, the dominant source is from the cosmic neutrino background (CNB), a relic from the early universe when it was about 1 second old, and has never been directly detected, although there are some proposed projects~\cite{cnb,Long:2014zva,Stodolsky:1974aq,Bauer:2022lri,Shergold:2021evs,Bauer:2021uyj,vogel,Huang:2016qmh,Zeldovich:1981wf,Shvartsman:1982sn}. On the other hand, for low-energy neutrinos at or below the MeV scale, due to their small de Broglie wavelength, it is possible for them to scatter off nuclei coherently, which has only recently been observed at the tens of MeV scale~\cite{COHERENT:2017ipa}.

Dark matter detection at or below MeV region is difficult and current limits are much weaker than at the GeV region. Traditional direct searches for dark matter, looking for nuclear recoils in deep underground detectors, are challenging due to insufficient recoil energy, and the experimental searches of cold sub-GeV dark matter have focused on the Migdal effect~\cite{migdal} and the interaction with electrons~\cite{EDELWEISS,SENSEI}. Low-mass dark matter searches with liquid Helium has also been proposed~\cite{Liao:2021npo,Liao:2022zqg}. Recently, experiments also exploit the fact that a fraction of the cold dark matter is boosted to relativistic energies and thus can be efficiently detected in direct detection experiments~\cite{Super-Kamiokande:2022ncz,PandaX-II:2021kai,Bringmann:2018cvk,Plestid:2020kdm,Hu:2016xas}. More details on low-threshold dark matter direct detection in the next decade can be found in Ref.~\cite{Essig:2022dfa}.

On the other hand, novel methods and rich phenomena are crucial to keep high-energy physics more robust and attractive~\cite{Lu:2022ibc}. Recently, atom techniques have developed quickly with rich applications in fundamental physics, for example, the MAGIS-100~\cite{MAGIS-100:2021etm} and AION~\cite{Badurina:2019hst} experiments propose to use atom interferometry to detect gravitation wave and ultra-light dark matter; the ALPHA experiment~\cite{ALPHA:2013skd} measured the gravitational mass of antihydrogen atoms; and the STE-QUEST proposal~\cite{Aguilera:2013uua} will test the Universality of free fall using cold atom interferometry.

\begin{figure}
    \centering
    \includegraphics[width=.9\columnwidth]{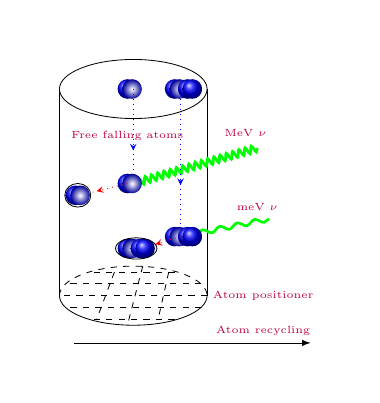}
    \caption{Illustration of experiments to detect low energy neutrinos and low mass dark matters with freely falling atoms. The resulting kinematic shifts of atom clusters are visible and can be measurable. At low energy or low mass, coherent scattering of neutrinos and dark matters with nuclei will be enhanced significantly.}
    \label{fig:eon}
\end{figure}

\section{Freely flying atom clusters}

In this paper, we are interested in a novel method to detect low-energy neutrinos with freely falling atoms, as shown in Fig.~\ref{fig:eon}. These atoms, if collided by surrounding dark matter or neutrinos from the cosmological background, from the Sun or other sources, may be shifted horizontally, creating a clear signal that can be measured. For $10^7$ heavy atoms of a mass number around 100 with a small recoil energy as 1 meV, the velocity can reach
\begin{equation}
v\sim \sqrt{2\times 1.6\times 10^{-22}/(1.7\times 10^{-18})} \sim 0.01\, {\rm m/s},
\end{equation}
which can produce significant kinematic shift that can be detected, while it would be almost impossible using conventional method.

\begin{figure}
    \centering
    \includegraphics[width=.8\columnwidth]{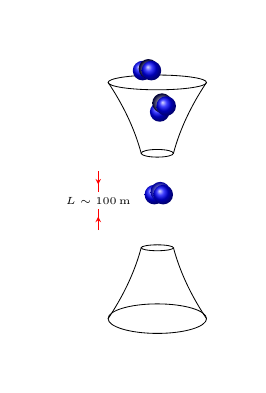}
    \caption{Illustration of atoms injection. The funnels are used to guided atoms drop down vertically. Technology challenges include high quality atom powder production and injection. Notice for sub MeV scale dark matter, the requirements are relatively loose: with $10^2$ atom guns shooting downwards CsI clusters consisting of $10^{7}$ atoms with a frequency around $10^3$ Hz, it can already be sensitive to scattering cross section at $10^{-33}\rm{cm}^{2}$ and surpass current limits.}
    \label{fig:eon2}
\end{figure}

The atoms can be dropped down in a form of a crystal-like cluster consisting of $N_d=10^{7\, (8)}$ or even more (see below). For meV neutrinos,  the de Broglie wavelength will be at mm level and thus can scatter with $N_d$ atoms coherently with a much enhanced cross section. Although the recoil energy will be shared by those atoms, resulting velocity may still lead to visible shifts to be measured by position detector, as shown in Fig.~\ref{fig:eon} and Fig.~\ref{fig:eon2}. The atoms dropped further down can be collected and recycled to be dropped down again.

\section{Low Energy Neutrino Probe}

We take two benchmarks as examples, i.e., 1 MeV and 50 meV neutrinos. The neutrino energy flux on Earth integrated over directions and summed over flavors, $F=E\phi$, at 1 MeV (50 meV) scale is around $10^{11\,(9)}\rm{cm}^{-2}\rm{s}^{-1}$ ~\cite{Vitagliano:2019yzm} (notice also CNB flux can be larger due to gravitational clustering~\cite{vogel}). Neutrino nuclei scattering cross section $\sigma\sim 10^{-43 (-58)}\rm{cm}^{2}$ ~\cite{Formaggio:2012cpf,Hagmann:1998nz,Yanagisawa2014}, for 1 MeV (50 meV) neutrinos.
%http://cupp.oulu.fi/neutrino/nd-cross.html

For 1 MeV (50 meV) neutrinos,  the de Broglie wavelength of the neutrino nuclei scattering process, $\lambda_\nu$, is about 1243 fm (4 microns)! The rates of coherent elastic neutrino-nucleus scattering will then be enhanced by about neutron number squared $N^2\sim 10^{4}$ for CsI as an example (notice the atomic numbers of the Cs and I nucleus are similar).  Notice that for very low energy neutrinos at KeV or even meV energy scale, the rates of coherent elastic neutrino-nucleus scattering will not only be enhanced by neutron number squared $N^2$~\cite{Formaggio:2012cpf}, but may also be further enlarged by a factor of $N^2_c$,  due to the even larger $\lambda_\nu$ than the target size, where $N_c=(N_{Av}/A)\rho(\lambda_\nu)^3$, with $N_{Av}$ as the Avogadro constant $\sim 6.02\times 10^{23}$ and the target density $\rho=$ 4.5 $\rm{g}/\rm{cm}^3$ for CsI. 

For MeV neutrinos, we have $N_c\sim 1$, while for neutrinos at 50 meV energy scale, the enhance factor $N_c \sim 10^{12}$, opening the possibility of neutrino scattering with vast numbers of atoms. However, it is not realistic to drop huge amount of atoms in a single droplet, as the recoil velocity will be tiny. Instead, it would be more useful to drop around, for example, $N_d\sim 10^{8}$ atoms' cluster. Correspondingly, the rates of coherent elastic neutrino-nucleus scattering will be enhanced in total by $N^2\times N_c^2\sim 10^{20}$ with $N_c=N_d$ here. 

In our proposal, atom clusters are kept dropping freely down from the top. For a vertical height of $L\sim 100$ meters, the flying time is around 5 seconds.  To keep the total amount of atoms around $N_{cl.}\sim 10^{22}$ (i.e., Avogadro number like) at any time within the cylinder, $10^{28}$ atoms need to be dropped in one year, i.e., 1 tons of CsI, of which mostly can be recycled. Moreover, as we will see below, the requirement for low mass dark matter search will be much relaxed.

For neutrinos scattered coherently with a stable cluster of $N_d$ atoms, the maximum recoil energy is ~\cite{vogel, Vergados:2008jp, Formaggio:2012cpf}:
\begin{equation}
T_{\rm max} = \frac{E_\nu}{1 + (N_d\times {\rm M_{atom}})/(2E_\nu)}.
\end{equation}

For the case of MeV neutrinos with $N_d\sim 10^{8}$, the maximum velocity is around 0.1 mm/s, the horizontal maximum shift distance will be around 1 mm.  However, it may be more practical to drop larger CsI clusters consisting of $N_d\sim 10^{11}$ atoms. In that scenario, the maximum shift velocity is around $10^{-7}$ m/s, and the horizontal maximum shift distance will be around 1 micron. 

For the case of 50 meV neutrinos with $N_d\sim 10^{7(8)}$,  the resulting maximum velocity is around $10^{-11(-12)}$m/s. %We can estimate the transferred momenta $p=E/c\sim 2\times 10^{-29}$ kg.m/s. Thus the %atoms will in average have a velocity as $p/A/m_p/N_d\sim 3\times 10^{-8}$m/s.
On the other hand, large fraction of such low energy neutrinos may be non-relativistic with a velocity as 0.1 c. The maximum recoil energy is~\cite{Bringmann:2018cvk}
\begin{equation} \label{nonr}
E^{\rm max}_{\rm recoil} = \frac{(2\times {\rm M_D} \times v)^2}{N_d \times {\rm M_{atom}}}, 
\end{equation}
and the resulting recoil velocity can be one magnitude lower than the above relativistic case.  

Correspondingly, after falling a distance of 100 m into the bottom of the device, the horizontal maximum shift distance will be around 0.1 nm, which should still be measurable (with e.g. laser method) although very challenging. Further optimization can be possible by adjusting atom dropping cluster size and vertical height.

Summing over all factors, we can estimate the neutrino atom(s) scattering rate per second as 
\begin{equation}
N \sim F\times (N^2\times N^2_c \times \sigma) \times N_{cl.} ,
\end{equation}
For 1 MeV and 50 meV neutrinos, it reads about $10^{11}\times10^{-39}\times 10^{22}\sim 10^{-6}$ and $10^{9}\times10^{-38}\times 10^{23}\sim 10^{-6}$  per second, i.e., 10 events per year for each, respectively. Similarly, for 0.1 MeV, the expected event yield can reach 1-10 per year.  

Thus our proposal can be useful for measuring coherent neutrino scatterings at 0.1 -1 MeV region to probe solar neutrinos physics.  With $10^4$ atom guns shooting downwards CsI clusters consisting of $10^{11}$ atoms, with an initial velocity of 1-10 m/s and time gap around $10^{-6}$ per second. One can expect 10 events to be observed every year.  

On the other hand, although it is still challenging to touch CNB, one can expect to get closer and closer to the floor by improving atom dropping rates. The device should be able to probe vast regions of very low energy neutrinos, for the first time. 

\section{Low mass dark matter searches}

It also has great potential to detect low mass dark matter.  The local density of dark matter is at the order of 0.3 GeV/$\rm{cm}^3$ and with typical velocity of $v=300$\, km/s~\cite{Benito:2019ngh,ParticleDataGroup:2022pth}. The dark matter flux then can be estimated as $10^7/\rm{M_D}[\rm{GeV}]\rm{cm}^{-2}\rm{s}^{-1}$. When the dark matter mass $\rm{M_D}\sim$1 (0.1) MeV, the flux will be  $10^{10\,(11)}\rm{cm}^{-2}\rm{s}^{-1}$.  Similarly as above for neutrino studies, we can estimate the dark matter atom(s) scattering rate per second as 
\begin{equation}
10^{10\, (11)}\times (10^4 \times \sigma_D) \times N_{cl.}.
\end{equation}
With $100$ atom guns shooting downwards CsI clusters consisting of $N_d=10^{7}$ atoms with a frequency of $10^{3}$ Hz. For this case, the total amount of atoms at any time within the cylinder is around $N_{cl.}\sim 10^{12}$ and much smaller than the above mentioned neutrino probe case. 

The resulting maximum velocity following Eq.~\ref{nonr} is around $10^{-7}$ m/s, the horizontal maximum shift distance will be around 1 micron. Within one year, the sensitivity will be $\sigma_D\sim 10^{-33 (-34)}\rm{cm}^{2}$ for $\rm{M_D}=1 (0.1)$ MeV, and already surpass current limits~\cite{CRESST:2019jnq,Super-Kamiokande:2022ncz,PandaX-II:2021kai}, yet without the assumption of Cosmic Ray Boosted mechanism~\cite{Super-Kamiokande:2022ncz,PandaX-II:2021kai,Bringmann:2018cvk}. 

\section{Technique challenges and opportunities}

For the moment, we only consider horizontal shift. In realistic, however, neutrinos and dark matters can fly in from any directions, a 3D like position detector may be more powerful.

Technology challenges in the proposal here also include high quality atom powder production and high frequency injection. Experiences from atom or condensed matter physics community will be extremely beneficial. For example, 
there have been extensive studies~\cite{atomclusters} on properties and productions of clusters with hundreds to thousands, or millions to billions of atoms. On the other hand, we believe that this proposal may also stimulate interdisciplinary development among high energy physics and other communities.

\section{Outlook and conclusion} 
We propose a new method to detect low-energy neutrinos and low-mass dark matter at or below the MeV scale, through their coherent scatterings from freely falling heavy atoms and the resulting kinematic shifts. We start with a simple calculation for illustration: for $10^7$ heavy atoms of a mass number around 100 with a small recoil energy of 1 meV, the corresponding velocities can reach $0.01, {\rm m/s}$ and produce significant kinematic shifts that can be detected. We then show that the proposed device should be able to probe vast low-energy regions of neutrinos from meV to MeV and can surpass previous limits on sub-MeV dark matter by several orders of magnitude. Such a proposal can be useful to (1) detect sub-MeV-scale dark matter: with $10^2$ atom guns shooting downwards, for example, CsI or lead clusters consisting of $10^{7}$ atoms with a frequency around $10^3$ Hz, it can already be sensitive to scattering cross-sections at the level of $10^{-33 (-34)}\rm{cm}^{2}$ for 1 (0.1) MeV dark matter and surpass current limits. Technological challenges include high-quality atom cluster production and injections. (2) Measure coherent neutrino-nuclei scatterings at the 0.1-1 MeV region for the first time: with $10^4$ atom guns shooting downwards CsI clusters consisting of $10^{11}$ atoms and a frequency of $10^{6}$ Hz. One can expect 10 events from MeV solar neutrinos to be observed per year. Furthermore, (3) this method can be extended to probe very low-energy neutrinos down to the eV-KeV region and may be able to detect the cosmic neutrino background, although it remains challenging.

This kind of falling atom device should be quite flexible and robust to probe very low energy neutrino down to eV-KeV region. By optimizing over the atom type and sizes, it could be sensitive to neutrinos and dark matters covering vast range of energy or mass scales.

\appendix
%\section{}
\begin{acknowledgments}
This work is supported in part by the National Natural Science Foundation of China under Grants No. 12150005, No. 12075004 and No. 12061141002, by MOST under grant No. 2018YFA0403900.
\end{acknowledgments}

\bibliographystyle{ieeetr}
\bibliography{h}
\end{document}